\documentclass[prl,showpacs,twocolumn]{revtex4}

\usepackage{graphicx}

\begin{document}

\title{The Larmor clock and anomalous spin dephasing in silicon}

\author{Biqin Huang}
\altaffiliation{Present address: HRL Laboratories, Malibu CA}
\author{Ian Appelbaum}
\altaffiliation{appelbaum@physics.umd.edu}
\affiliation{Center for Nanophysics and Advanced Materials and Department of Physics, University of Maryland, College Park MD 20742 USA}

\begin{abstract}
Drift-diffusion theory -- which fully describes charge transport in semiconductors -- is also universally used to model transport of spin-polarized electrons in the presence of longitudinal electric fields. By transforming spin transit time into spin orientation with precession (a technique called the ``Larmor clock'') in current-sensing vertical-transport intrinsic Si devices, we show that spin diffusion (and concomitant spin dephasing) can be greatly enhanced with respect to charge diffusion, in direct contrast to predictions of spin Coulomb-drag diffusion suppression.
\end{abstract}


\maketitle

Observation of spin precession and dephasing is not only clear evidence for nonequilibrium spin injection into nonmagnetic materials\cite{JOHNSON, MONZON}, but it also provides a means to quantitatively measure electron transport dynamics.\cite{APS} As shown recently, spin precession measurements in a perpendicular magnetic field can be used to directly derive the empirical spin transit time distribution via the Fourier transform.\cite{LATERAL} This ability to extract time-of-flight data from a quasistatic measurement is enabled by the encoding of spin transit time $t$ into spin precession angle $\theta=\omega t$ (where $\omega$ is spin precession frequency), a technique called the ``Larmor clock'' in the context of transport theory\cite{BAZ, BUTTIKER, HARTMAN, SHERMANEPL}.  
 
The power of this method to reveal the details of spin transport is diminished, however, by systematic sources of spin ``dephasing'' (uncertainty in spin orientation, $\Delta\theta$). In lateral voltage-sensing\cite{LOU, JONKERLATERAL, SHIRAISHIHANLE, VANWEESGRAPHENE} or current-sensing\cite{2MM, LATERAL} devices, for example, there is strong geometric spin dephasing due to transit-length uncertainty induced by nonzero injector and detector width along the transport channel dimension. Vertical-transport devices, on the other hand, have true transverse symmetry, eliminating this systematic geometric spin dephasing and allowing the transport distribution to expose the consequences of the fundamental transport phenomena unencumbered by spurious effects\cite{APPELBAUMNATURE, BIQINPRL, DEPHASING}. 

In this Letter, we use the concept of the Larmor clock to identify a phenomenon of anomalous spin diffusion, which behaves in a manner strikingly different from both the expectations of charge transport in the nondegenerate regime and our present understanding of spin transport in semiconductors. Several potential physical origins are discussed, highlighting the need for a more sophisticated drift-diffusion theory necessary to accurately model nonequilibrium spin transport in nonmagnetic materials.  

\begin{figure}
\includegraphics[scale=0.425]{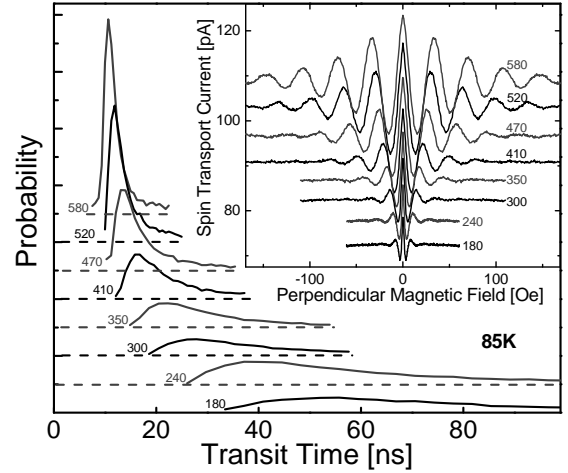}
\caption{Transit-time distributions for spin current traveling through 350$\mu$m undoped Si under different internal electric fields (180-580 V/cm) at a temperature of 85K. Inset: Symmetrized spin precession data, from which these distributions were calculated using the Fourier transform method described in Ref. \onlinecite{LATERAL}.
\label{FIG1}}
\end{figure}

Fig. \ref{FIG1} shows an example of spin current transit-time distribution extraction at a fixed temperature $T=$85K for different internal electric fields $E$=180 - 580 V/cm in vertical-transport $L=$350$\mu$m-thick Si(100) devices utilizing ballistic hot electron injection and detection\cite{APPELBAUMNATURE, BIQINPRL, BIQINJAP}(For a review of the device details and related techniques, see Ref. \onlinecite{APPELBAUMREVIEW}.) The inset shows raw spin precession data, symmetrized to avoid the effects of ferromagnetic contact switching and maintain a real-valued Fourier transform. We control the spin precession frequency through the transverse magnetic field $B$, via $\omega=g\mu_B B/\hbar$, where $g$ is the electron g-factor, $\mu_B$ is the Bohr magneton, and $\hbar$ is the reduced Planck constant. Oscillations as a function of magnetic field with average period $2\pi\hbar/g\mu_B \bar{t}$, where $\bar{t}$ is average transit time\cite{BIQINJAP}, indicate coherent spin precession. The amplitude suppression of these oscillations at higher magnetic field is the direct result of spin dephasing from nonzero $\Delta\theta$ due to spin diffusion. As expected from basic charge transport, higher drift electric fields result in shorter average transit times and lower transit time uncertainty. These measurements can be repeated at different temperatures, and the normalized distributions obtained ($P(t)$) can be analyzed to determine any distribution moment. 

\begin{figure}
\includegraphics[scale=0.45]{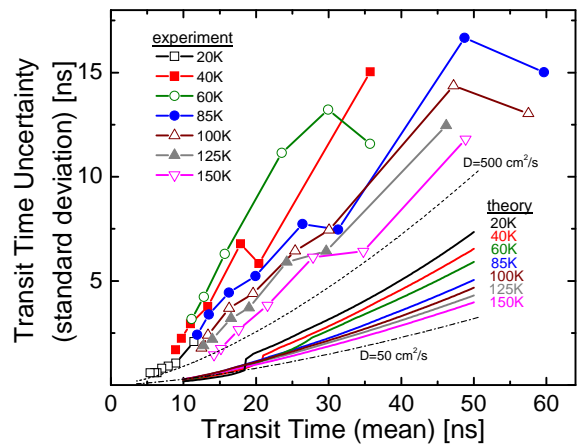}
\caption{Comparison of experimental and theory spin current transit time uncertainty (standard deviation of transit time distributions). Larger uncertainty corresponds to higher spin dephasing. Theory results rely on the drift-diffusion Green's function (Eqn. \ref{GREENSFUNC}), with Canali et al's data in Ref. \onlinecite{CANALI} and the nondegenerate Einstein relation for accurate diffusion coefficient modeling.  
\label{FIG2}}
\end{figure}

Most important here is the transit-time uncertainty (distribution standard deviation $\Delta t=\sqrt{\int^\infty_0P(t)(\bar{t}-t)^2dt}$), which is a direct measure of spin dephasing through $\Delta\theta=\omega\Delta t$. This quantity is plotted in Fig. \ref{FIG2}, along with similar data measured at other temperatures between 20K and 150K. All constant-temperature data show a similar trend of increasing uncertainty as the mean increases in smaller drift fields. 

The model universally used to simulate spin precession measurements like these incorporates only drift (due to an electric field $E$) at velocity $v$, diffusion (caused by random thermal motion) with coefficient $D$, and spin relaxation with timescale $\tau$. The resulting Fokker-Planck rate equation is nominally identical to the minority-carrier equation used to successfully model charge transport in semiconductors,\cite{HAYNES} except that it models the density of a single component of the spin vector rather than charge density. Due to the absorbing boundary conditions relevant for our current-sensing devices, the transit-time distribution determined from the Green's function $s(x,t)$ of this scalar drift-diffusion-relaxation equation is\cite{BCS}

\begin{equation}
P(t)=s(x=L,t)=\frac{L}{t}\frac{1}{2\sqrt{\pi Dt}}e^{-\frac{(L-vt)^2}{4Dt}}e^{-t/\tau}.
\label{GREENSFUNC}
\end{equation}

We expect our empirically measured distributions to have this form, for some value of $D$, $v$, and $\tau$. Spin lifetime $\tau$ has been measured in undoped Si\cite{BIQINPRL} and its temperature dependence has been analyzed theoretically\cite{FABIANWU}; typical values are much longer than the transport times measured ($<60ns$ in Fig. \ref{FIG2}) and so have negligible effect on the transit-time uncertainty. Because electron density is in the nondegenerate regime in our experiments, it may appear reasonable to apply the fluctuation-dissipation theorem (Einstein relation, $D=\mu k_BT/q$, where $k_B$ is the Boltzmann constant and $q$ is the fundamental electron charge) which provides a relationship between $D$ and $v$ via the mobility $\mu=v/E$. 

Using this simple theory, we can generate predictions of spin dephasing at constant $D$, as shown by the dot-dash and dashed lines (corresponding to 50cm$^2$/s and 500cm$^2$/s, respectively) in Fig. \ref{FIG2}. It is clear that the empirical data shown (labeled ``experiment'') have effective diffusion coefficients in excess of the latter, larger, value. When the Einstein relation (along with well-established charge transport data from Ref. \cite{CANALI}) is used, uncertainties corresponding to diffusion coefficients of much lower magnitude (labeled ``theory'' in Fig. \ref{FIG2}) are predicted.   

What is the cause of this experimentally-observed enhancement of spin dephasing? It has been known for some time that spin Coulomb drag\cite{VIGNALE, SCDEXPT} causes an inequality between charge diffusion and spin diffusion coefficients and inapplicability of the Einstein relation, but always {\it suppressing} spin diffusion with respect to charge diffusion. Here, however, we see an apparently large enhancement. Because the Larmor clock depends on a well-defined precession frequency $\omega$, uncertainties in this quantity $\Delta\omega$, caused by inhomogeneous magnetic fields or g-factor variations, can certainly contribute to dephasing. However, we have confirmed that the drastic spin diffusion enhancement seen here would require inhomogeneities of approximately 50\%, which is inconsistent with our experimental conditions.

\begin{figure}
\includegraphics[scale=0.6]{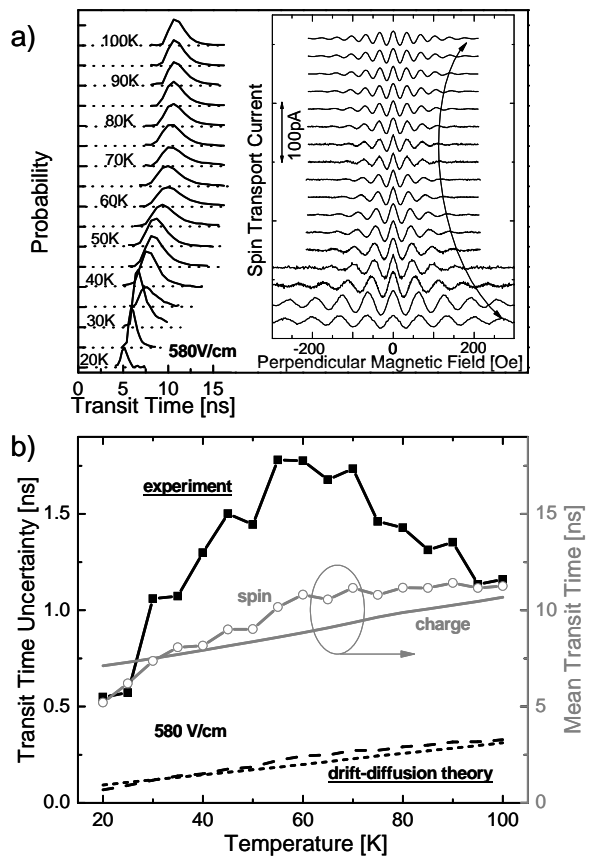}
\caption{Temperature dependence of spin dephasing in undoped Si showing (a) spin current transit time distributions obtained from spin precession measurements (shown in inset, where the curved arrow emphasizes the extent of the envelope function modulating the coherent oscillations) in an internal electric field of 580 V/cm. (b) Temperature dependence of transit time uncertainty (standard deviation of the distributions) showing nonmonotonic behavior that contrasts with the monotonic increase expected from drift-diffusion theory (Eqn. \ref{GREENSFUNC}) using the Einstein relation. Long dashes use the empirical mobility, whereas short dashes use charge transport data from Ref. \cite{CANALI}. Grey data (and right axis) shows the mean transit time of the empirical distribution (open circles, ``spin'') and prediction from Ref. \cite{CANALI} (solid line, ``charge'').
\label{FIG3}}
\end{figure}

Even more striking evidence for anomalous dephasing is presented in Fig. \ref{FIG3}, where we show temperature dependence at a fixed electric field of $E$=580 V/cm. The transit-time distributions are shown in Fig. \ref{FIG3}(a), along with the corresponding symmetrized spin precession data in the inset.  The spin dephasing temperature dependence (distribution standard deviation) is shown in Fig. \ref{FIG3}(b). Dephasing is small at the lowest temperature (20K), but rises to a maximum near 60K, after which it slowly falls as temperature increases toward 100K. Even without calculating the standard deviation from corresponding transit-time distributions, it is clear from the raw data in the inset of Fig. \ref{FIG3}(a) that the width of the envelope function of the precession oscillations -- inversely proportional to the dephasing -- has a minimum near 60K. This behavior cannot be explained within the drift-diffusion model using the fluctuation-dissipation theorem, which predicts a monotonic increase in dephasing as a function of temperature.\cite{DEPHASING} This prediction is shown using both the empirical mobility ($\mu=L/\bar{t}E$) and charge-transport data (again, from Ref. \onlinecite{CANALI}) as dashed and dotted lines, respectively; they are nearly identical. Once again, in addition to the monotonic nature of this prediction, the magnitude is far below what is seen experimentally. Systematic sources of dephasing are temperature {\it independent}, so the phenomenon seen here is clearly intrinsic to the details of spin precession during transport in Si.
 
This nonmonotonic behavior is somewhat suggestive of thermally-activated electron trapping during transport. In such a model, average electron dwell times ($\tau_t$) before emission from shallow traps are exponentially dependent on temperature according to the Arrhenius law: at high temperature, the trap time is very short, so electron trapping has negligible effect on the transit time distribution compared with the diffusion-induced distribution width. At low temperatures, the trapping time can be much longer than the spin lifetime, so when a trapped electron is eventually released, it has become depolarized and makes no contribution to the overall spin signal. However, at intermediate temperatures, the effects of electron trapping on distribution broadening may be significant. 

Let us now evaluate the applicability of this model. Due to the unimodal form of the distributions at intermediate temperatures shown in Fig. \ref{FIG3}(a), nearly all of the electrons would have to be affected by such traps, so the expected number of average trapping events $k$ is large. In this limit, the convolution of $k$ exponential-decay trap escape distributions (the Erlang distribution $G(t;k,\tau_t)$) approaches a gaussian form with mean total trapping time $k\tau_t$ and standard deviation $\sqrt{k}\tau_t$\cite{BELLERLANG}. Therefore, if this trapping model is applicable, the deviation of the mean spin transit time from the expected charge transport time should be more than the difference in the observed standard deviation from that predicted using charge transport data for any $k>1$. However, as shown by the mean transit time data (grey, right axis) in Fig. \ref{FIG3}(b), this is not the case; spin and charge mobility are nearly identical.

\begin{figure}
\includegraphics[scale=0.4]{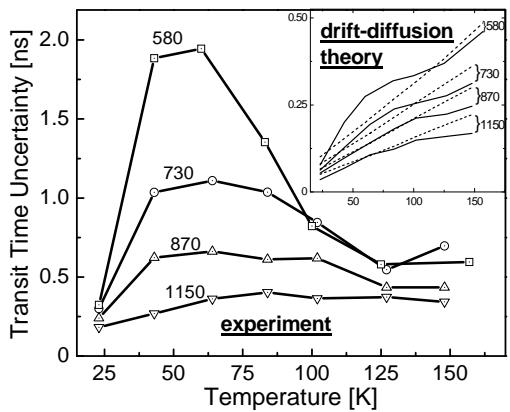}
\caption{Suppression of anomalous spin dephasing with electric field from 580 V/cm to 1150V/cm.  Inset: Drift-diffusion predictions using the Einstein relation. Solid lines use the empirical mobility, whereas dashed lines use charge transport data from Ref. \cite{CANALI}. 
\label{FIG4}}
\end{figure}

Furthermore, as shown in Fig. \ref{FIG4}, increasing the internal electric field only a nominal amount from 580V/cm to 1150V/cm strongly suppresses this temperature-dependent anomalous spin dephasing far beyond what is expected from trap ionization.  Shallow trapping is thus not the cause of our observations of enhanced spin diffusion. The observed anomalous dephasing may therefore be regarded as the result of a more fundamental, intrinsic, cause. For instance, due to the direct relationship between electron density and transit time in our insulating samples\cite{DEPHASING}, higher electric fields result in more dilute electron densities and modify the electron-electron interaction which may play a role in spin dephasing.

Because spin is a vector quantity, modeling transport as a scalar as is typically done in the drift-diffusion equation is necessarily incomplete. A more complete, three-component vector drift-diffusion theory would account for this, including tensor representation of the spin diffusion coefficient.\cite{STANESCU} It remains to be seen, however, whether the anomalous behavior described here would be captured solely by this more sophisticated framework. 

In addition, the interaction of phonons with spin-polarized electrons may become relevant at the transport lengthscale investigated here. In light of recent theoretical developments with sophisticated calculations of the effects of spin polarized electron-phonon interactions utilizing detailed bandstructure\cite{FABIANWU, DERYSI}, it is suggestive that thermal conductivity due to three-phonon Umklapp processes also show nonmonotonic behavior\cite{HOLLAND} with a peak near 60K as in Fig. \ref{FIG3}(b). Although study of the Elliott-Yafet electron-phonon interaction has focused solely on spin relaxation\cite{YAFET, FABIANWU}, it may be that similar physics has an effect on spin dephasing when all three components of spin are taken into account in a transport scenario. 

Finally, we wish to point out that this anomalous dephasing behavior may be a universal characteristic of nonequilibrium spin transport processes. A notable example is evident in optical measurements of the persistent spin helix in two-dimensional GaAs\cite{SPINHELIX}, where nonmonotonic temperature behavior of the spin-orbit-enhanced mode lifetime is observed. Because the ``persistence'' of the spin helix relies on control of spin diffusion, the similarity in behavior with what we present here in Figs. \ref{FIG3} and \ref{FIG4} is suggestive of analogous underlying physics despite very different materials and even dimensionality.\cite{WU3, WU4}     

In conclusion, we have shown how the Larmor clock concept can be used as a powerful probe of spin dynamics in Si by providing a means to encode fast time-dependent transport processes into spin orientation that is measured quasistatically. A drastic enhancement of spin dephasing seen especially in nonmonotonic temperature dependence (from devices where geometric dephasing is eliminated by transverse symmetry) cannot be explained by our present understanding of spin transport which would contrarily predict a suppression of spin dephasing due to spin Coulomb drag. The challenges to construction of a successful theory capturing the observed phenomenon have been discussed; it is hoped that the suggestions here provide a theoretical starting point for a more sophisticated model of nonequilibrium spin transport in nonmagnetic semiconductors. 

\acknowledgments{
We gratefully acknowledge helpful discussions on the trapping model with B. Kane and S. Bohacek, on the vector drift-diffusion theory and spin helix with V. Galitski, and on umklapp processes in Si with W. Wulfheckel.  This work was funded by the Office of Naval Research and the National Science Foundation.
}

\end{document}